\documentclass[aps,prd,nofootinbib,superscriptaddress,notitlepage,preprintnumbers,twocolumn,reprint]{revtex4-1}

\usepackage{graphicx}
\usepackage{amsmath,graphicx,verbatim,epsfig}
\usepackage{epstopdf}
\usepackage[utf8x]{inputenc}
\usepackage[colorlinks=true,linkcolor=blue,citecolor=blue]{hyperref}
\usepackage[usenames,dvipsnames]{color}
\usepackage{bm}
\usepackage{cancel}
\usepackage[normalem]{ulem}

\newcommand{\eps}{\varepsilon}

\newcommand{\nn}{\nonumber}

\newcommand{\bn}{{\bar n}}

\newcommand{\be}{\begin{equation}}
\newcommand{\ee}{\end{equation}}
\newcommand{\bea}{\begin{eqnarray}}
\newcommand{\eea}{\end{eqnarray}}
\newcommand{\balign}{\begin{align}}
\newcommand{\ealign}{\end{align}}

\newcommand{\sandwich}[3]{\left< #1 \right | #2 \left | #3 \right >}

\newcommand{\bg}{\begin{gather}}
\newcommand{\foma}{\end{gather}}

\newcommand{\noopsort}[1]{}

\newcommand{\vecb}[1]{\mbox{\boldmath $#1$}}
\newcommand{\vecbe}[1]{\mbox{\boldmath ${\scriptstyle #1}$}}

\def\e{\epsilon}

\def\z{\zeta}

\def\<{\langle}
\def\>{\rangle}

\def\g{\gamma}  \def\G{\Gamma}
\def\d{\delta}

\def\m{\mu}
\def\n{\nu}
\def\t{\tau}

\def\z{\zeta}

\def\({\left(}
\def\[{\left[}
\def\){\right)}
\def\]{\right]}

\def\ln{\hbox{ln}}

\def \le { \left    }
\def \ri { \right }

\def\bP{\bar P}

\usepackage[normalem]{ulem} 
\renewcommand\sout{\bgroup \color[rgb]{1,0,0} \ULdepth=-.5ex \ULset}

\newcommand{\eq}[1]{Eq.~\eqref{#1}}

\begin{document}
\preprint{NIKHEF 2015-34}
\title{The Universal Transverse Momentum Dependent Soft Function at NNLO}

\author{Miguel G. Echevarria}
\email[]{m.g.echevarria@nikhef.nl}
 \affiliation{Nikhef Theory Group, Science Park 105, 1098XG Amsterdam, the Netherlands}
\affiliation{Department of Physics and Astronomy, VU University Amsterdam, De Boelelaan 1081, NL-1081 HV Amsterdam, the Netherlands}

\author{Ignazio Scimemi}
\email[]{ignazios@fis.ucm.es}
\affiliation{Departamento de F\'isica Te\'orica II, Facultad de Ciencias F\'isicas, Universidad Complutense de Madrid, Spain}

\author{Alexey Vladimirov}
\email[]{vladimirov.aleksey@gmail.com} \affiliation{Institut f\"ur Theoretische Physik, Universit\"at Regensburg, D-93040 Regensburg, Germany}

\date{\today}

\begin{abstract}

All (un)polarized transverse momentum dependent functions (TMDs), both distribution and fragmentation functions, are defined with the same universal soft function, which cancels spurious rapidity divergences within an individual TMD and renders them well-defined hadronic quantities.
Moreover, it is independent of the kinematics, whether it is Drell-Yan, deep inelastic scattering or $e^+e^-\rightarrow 2$ hadrons.
In this paper we provide this soft function at next-to-next-to-leading order (NNLO), necessary for the calculation of all TMDs at the same order, and to perform the resummation of large logarithms at next-to-next-to-next-to-leading-logarithmic accuracy.
From the results we obtain the $D$ function at NNLO, which governs the evolution of all TMDs. 
This work represents the first independent and direct calculation of this quantity.
Given the all order relation through a Casimir scaling between the soft function relevant for gluon TMDs and the one for quark TMDs, we also obtain the first at NNLO.
The used regularization method to deal with the rapidity divergences is discussed as well.

\end{abstract}

\maketitle

\section{Introduction}

In the recent works that establish the factorization theorems for the transverse momentum spectra in Drell-Yan (DY), Vector/Higgs boson
production, semi-inclusive deep inelastic scattering (SIDIS) and $e^+e^-\rightarrow 2$ hadrons~\cite{Collins:2011zzd,GarciaEchevarria:2011rb,Echevarria:2012js,Echevarria:2014rua}, the soft function (SF) plays a central role.
The main building blocks within these factorization theorems are the transverse momentum dependent functions (TMDs), which encode the relevant non-perturbative physics.
And the SF enters these TMDs in a particular way, in order to cancel spurious rapidity divergences and make them well-defined hadronic quantities.

The SF is defined as the vacuum expectation value of a product of light-like Wilson lines evaluated at an space-like distance.
It is an utterly singular object which is undefined without a proper regularization method, and cannot be considered as a probability amplitude in the usual quantum field theory sense.
Given the way it is introduced and used, the SF is similar to the renormalization constants of quantum field theory ($Z$ factors), which are undefined without regularization, and has individual physical meaning: while the renormalization constants subtract the ultraviolet (UV) divergences of diagrams, the SF subtracts the rapidity divergences, the most specific divergences of TMD factorization.

When calculating the SF one faces divergences of different types and only some of them can be controlled by dimensional regularization (with $d\neq4$).
Therefore, it is important to understand the origin of these divergences.

The main attention is paid to the rapidity divergences, as they represent the most crucial point in the calculation of the SF.
These divergences arise when loop momentum, denoted by $k$, is boosted in one light-cone direction (given by the vectors $n_\pm$ with $n_+^2=n_-^2=0$ and $n_+n_-=2$), say $k^+\rightarrow\infty$ and $k^-\rightarrow 0$, keeping $k^+k^-$ fixed.
The important point is that light-like Wilson lines, and therefore the SF, are invariant under the coordinate rescaling in their own light-like directions.
This invariance leads to an ambiguity in the definition of rapidity divergences.
Indeed, the boost $k^+\to a k^+$, $k^-\to k^-/a$ (with $a$ an arbitrary number) leaves the SF invariant, while in the limit $a\to\infty$ reproduces the rapidity divergent configuration.
Therefore, without a regularization which breaks the boost invariance, the SF cannot be explicitly calculated.
In this sense, rapidity divergences cannot be regularized by any type of dimensional-like regularization.

There are many ways to break boost invariance and make the rapidity divergences manifest.
One way is to tilt the Wilson lines from the light-cone, such that $n_\pm^2>0$ and infinitesimal.
This procedure implies the use of different soft factors to deal with the self-energies of Wilson lines and avoid double counting, see e.g.~\cite{Collins:2011zzd,Aybat:2011zv,Bacchetta:2013pqa}.
In these works one combines the unevaluated integrands of collinear and soft matrix elements together to form integrals free from rapidity divergences. 
Implementing these combinations at two loops can be  demanding.
From the calculational point of view, we find much more economical to have a sufficient regularization for the  TMDs, such that every diagram gives a finite result. 
It also grants a higher universality of the obtained expressions, since many diagrams and even parts of calculations appear in different objects and can be reused. 
The SF is a good example of such universality, as once it is evaluated at NNLO, it can be used to calculate all leading-twist TMDs at the same order. 
In this respect, the formulation of the factorization theorem of~\cite{GarciaEchevarria:2011rb} is more suitable for a NNLO evaluation and we use it in the definitions of this paper.

Given the importance that the SF has per se for the establishment of the factorization theorem, we provide its explicit NNLO calculation in the present work.
The SF is defined as
\begin{widetext}
\begin{align} \label{eq:softf}
S(\vecb k_{s\perp}) &=
\int\frac{d^2\vecb b_\perp}{(2\pi)^2} e^{i\vecbe b _\perp \cdot \vecbe k _{s\perp}}
\frac{{\rm Tr}_c}{N_c}
\sandwich{0}{\le[S_n^{T\dagger} \tilde S_\bn^T \ri](0^+,0^-,\vecb b_\perp)
\le[\tilde S^{T\dagger}_\bn S_n^T\ri](0)}{0},
\end{align}
\end{widetext}
where $S_n$ and ${\tilde S}_\bn$ stand for soft Wilson lines along the light-cone directions $n$ and $\bn$ ($n^2=\bn^2=0,\; n\cdot\bn=2$).
The superscript $T$ on Wilson lines in~\eq{eq:softf} implies subsidiary transverse gauge links from the light-cone infinities to transverse infinity (see more details in~\cite{Belitsky:2002sm,Idilbi:2010im,GarciaEchevarria:2011md}).
These links guarantee gauge invariance and are necessary for calculations in singular gauges.
The present calculation has been performed in Feynman gauge, where the contribution of transverse links vanishes.

The establishment of the factorization theorem to all orders in perturbation theory relies on particular properties of the SF with respect to the rapidity regulator.
Essentially, the logarithm of the SF is at most linear in the logarithms generated by the rapidity divergences (using the $\d$-regularization, to be introduced in~\eq{eq:reg} in the next Section).
This guaranties that the SF can be factorized in two pieces \cite{Echevarria:2012js}, and in turn it allows to define the individual TMDs.
With the $\delta$-regularization this important relation reads
\begin{align}\label{eq:splitting}&
\tilde{S}({\mathbf L}_\m,{\mathbf L}_{\sqrt{\d^+\d^-}}) =
\tilde{S}^\frac{1}{2}({\mathbf L}_\m,{\mathbf L}_{\d^+/\nu})\,
\tilde{S}^\frac{1}{2}({\mathbf L}_\m,{\mathbf L}_{\nu\d^-})
\,,
\end{align}
where tildes mark quantities calculated in coordinate space, $\nu$ is an arbitrary and positive real number that transforms as $p^+$ under boosts and we introduce the convenient notation
$${\mathbf L}_{X}\equiv\ln(X^2 \vecb b^2 e^{2\gamma_E}/4).$$
Note that the relation in \eq{eq:splitting} is exact, i.e. it is valid to all orders in perturbation theory, as well as to all orders in the $\epsilon$-expansion.
To the best of our knowledge there is no general proof of this statement.
Consequently, one of the goals of our calculation is to establish the relation in \eq{eq:splitting} at NNLO, and thus verify the TMD factorization theorem at the same order.

The fact that one has a unique SF for different processes (with a different composition of initial and/or final states) is a direct consequence of the linear dependence on the logarithms of the rapidity divergences.
Moreover, one can extract the TMD evolution function $D$ \cite{Echevarria:2012js} from the SF (see Sec.~\ref{sec:SFp}).
In this paper we obtain it explicitly at NNLO and establish the Casimir scaling between the $D$ function relevant for quark TMDs and for gluon TMDs, which is valid to all orders in perturbation theory.
The realization of the cancellation of rapidity divergences at NNLO within one single TMD has been explicitly checked for the first time in~\cite{Echevarria:2015usa} for the case of
the unpolarized non-singlet fragmentation functions, and its generalization for the whole set of unpolarized TMDs is in preparation~\cite{ves}.

In the literature we have found that an early calculation of the SF was done in~\cite{Belitsky:1998tc} using just dimensional regularization, which is not appropriate to deal with rapidity divergences.
Before the development of the current TMD formalism, the universality of soft radiation was discussed in~\cite{Collins:2004nx} at one-loop level.
In the context of TMDs, the first soft function appeared in the work of Ji-Ma-Yuan~\cite{Ji:2004wu,Ji:2004xq}.
In that work it was argued that the SF in coordinate space should depend just on the transverse coordinate.
However, the regularization of the SF was done using tilted Wilson lines, and properly defining the TMDs by using that regulator requires the combination of several types of soft functions, as later discussed in~\cite{Collins:2011zzd}.
Moreover Ji-Ma-Yuan proposed the subtraction of the whole SF in the definition of the TMDs, which does not provide a complete cancellation of all rapidity divergences within one TMD.
Collins argued in~\cite{Collins:2011zzd} that in order to properly define one single TMD it is necessary to combine three soft functions in coordinate space and still use tilted Wilson lines.
The computation of (three) SF using Wilson lines off-the-light cone turns out to be not practical at higher orders in perturbation theory.
The calculation of the SF on-the-light-cone has been attempted by several authors and it is not free from difficulties~\cite{Becher:2010tm,Li:2011zp,GarciaEchevarria:2011rb,Echevarria:2012js,Chiu:2012ir}.
In~\cite{Becher:2010tm}, for instance, the integrals of the soft function are all scaleless to all orders in perturbation theory, so that the regulator that they propose is not suitable for establishing the TMD formalism; however it appears to be very efficient for the calculation of the total cross section.
The calculation of the DY cross section done at two loops in~\cite{Gehrmann:2012ze,Gehrmann:2014yya} in this sense is equivalent to the QCD calculation performed in~\cite{Catani:2012qa}.
In~\cite{Chiu:2012ir} the rapidity divergences are regularized by explicitly breaking Lorentz invariance through the introduction of a rapidity regulator.
A two-loop calculation of the SF using this regulator has not been performed yet to our knowledge.
The regulator proposed here allows the calculation of a single TMD and has been used for the complete calculation of the non-singlet unpolarized TMD fragmentation function in~\cite{Echevarria:2015usa}, using the TMD factorization theorem in the EIS formalism~\cite{GarciaEchevarria:2011rb,Echevarria:2012js}.


The explicit form of Wilson lines together with the regularization method that we have used is detailed in Sec.~\ref{sec:def}.
The result of the calculation of the SF at NNLO and the related discussion appears in Sec.~\ref{sec:results}.
In Sec.~\ref{sec:SFp} we discuss the properties of the SF and in particular we argue that the universal evolution function $D$ for all TMDs can be extracted solely from the SF, both for gluons and quarks.
Finally we  conclude in Sec.~\ref{sec:conclusions}.
The technical details of our calculations are given in the appendices.

\section{Definitions and Regularization}
\label{sec:def}

The  SF is written  more easily in coordinate space, where it takes the form
\begin{eqnarray} \label{eq:SF_def}
&&\tilde S(\vecb b_{T})
\\ \nn
&&
=
\frac{{\rm Tr}_c}{N_c} \sandwich{0}{\, T\le[S_n^{T\dagger} \tilde S_\bn^T \ri](0^+,0^-,\vecb b_T) \bar
T\le[\tilde S^{T\dagger}_\bn S_n^T\ri](0)}{0}
\,,
\end{eqnarray}
where we explicitly denote the ordering of operators.
The Wilson lines are defined as
\begin{align} \label{eq:SF_def2}
&S_{n}^T = T_{n(\bn)} S_{n}\,,
\quad\quad\quad\quad
\tilde S_{\bn}^T = \tilde T_{n(\bn)} \tilde S_{\bn}\,,
\nn\\
&S_n (x) = P \exp \left[i g \int_{-\infty}^0 ds\, n \cdot A (x+s n)\right]\,,
\nn \\
&T_{n} (x) = P \exp \left[i g \int_{-\infty}^0 d\tau\, \vec l_\perp \cdot \vec A_{\perp} (\infty^+,0^-,\vec x_\perp+\vec l_\perp \tau)\right]\,,
\nn \\
&T_{\bn} (x) = P \exp \left[i g \int_{-\infty}^0 d\tau\, \vec l_\perp \cdot \vec A_{\perp} (0^+,\infty^-,\vec x_\perp+\vec l_\perp \tau)\right]\,,
\nn\\
&\tilde S_\bn (x) = P\exp\le[-ig\int_{0}^{\infty} ds\, \bn \cdot A(x+\bn s) \ri]\,,
\nn\\
&\tilde T_{n} (x) = P\exp\le[-ig\int_{0}^{\infty} d\t\, \vec l_\perp \cdot \vec A_{\perp}(\infty^+,0^-,\vec x_\perp+\vec l_\perp\t) \ri]\,,
\nn\\
&\tilde T_{\bn} (x) = P\exp\le[-ig\int_{0}^{\infty} d\t\, \vec l_\perp \cdot \vec A_{\perp}(0^+,\infty^-,\vec x_\perp+\vec l_\perp\t) \ri]
\,,
\end{align}
and transverse gauge links $T_{n(\bn)}$ appear for the gauge choice $n \cdot A=0$ (or $\bn \cdot A=0$), while the rest of the Wilson lines appearing in \eq{eq:softf} are obtained by exchanging $n \leftrightarrow \bn$ and path-ordering $P$ with anti-path-ordering $\bP$.
We notice that these definitions apply for SIDIS kinematics~\cite{Echevarria:2014rua}, but as we will show below, the soft function turns out to be universal.

The choice of the IR and rapidity regularization scheme is one of the key ingredients in the calculation of TMDs. In our work we choose to
regularize the rapidity divergences with the $\delta$-regularization, that has been already used for the same purpose by many authors (see
e.g.~\cite{Echevarria:2014rua,GarciaEchevarria:2011rb,Cherednikov:2008ua,Vladimirov:2014aja}). In its original definition,
$\delta$-regularization consists in setting the $i0$ prescription of eikonal propagators finite. 
However this definition appears to be inefficient for high-loop computations. 
In particular it breaks the non-abelian exponentiation property of Wilson lines, which is crucial for
TMD factorization to hold.

Thus, in order to provide a more efficient computation we regularize the UV and IR-soft (mass) divergences using standard dimensional
regularization with $D=4-2\eps$. Then for rapidity divergences we define a $\delta$-regularization scheme at the operator level, which
consists in modifying the definition of Wilson lines as
\begin{eqnarray}\label{eq:reg}
\tilde S_{\bar n}(0)&=&P\exp\left[ -i g \int_0^\infty d\sigma A_{+} (\sigma \bar n)\right]
\nn\\
&\rightarrow&
P\exp\left[ -i g \int_0^\infty d\sigma
A_{+} (\sigma \bar n)e^{-\d^{+}\sigma}\right]\nn
\,,\\
 S_{n}(0)&=&P\exp\left[ i g \int^0_{-\infty} d\sigma A_{-} (\sigma n)\right]
\nn\\
&\rightarrow&
P\exp\left[ i g \int^0_{-\infty} d\sigma
A_{-} (\sigma  n)e^{+\d^{-}\sigma}\right]
\,,
\end{eqnarray}
where $\delta^\pm \to 0^+$.
The modified operator supplies the non-abelian exponentiation property. To be explicit, at the level of Feynman
diagrams in momentum space, the modified expressions for the eikonal propagators are written as (e.g. absorbing gluons by the Wilson line
$[\infty^+,0]$)
\begin{eqnarray}&&
\frac{1}{(k_1^+-i0)(k_2^+-i0)...(k_n^+-i0)} \nn\\\label{eq:reg_propgators} && \to
\frac{1}{(k_1^+-i\delta^+)(k_2^+-2i\delta^+)...(k_n^+-ni\delta^+)} \,,
\end{eqnarray}
where the gluons are ordered from infinity to zero (i.e. $k_n$ is the gluon closest to zero).
As a consequence of the rescaling invariance of the
Wilson lines (that is now explicitly broken by the parameters $\delta^\pm$), the expressions for diagrams in the SF depend on the single variable
$2\delta^+\delta^-/(n\bar n)=\delta^+\delta^-$.

The ordering of poles in the eikonal propagators \eq{eq:reg_propgators} is crucial for perturbative exponentiation with usual properties, such as non-abelian exponentiation theorem on color-factors \cite{Gatheral:1983cz,Frenkel:1984pz} or logarithmical counting
\cite{Sterman:1981jc}.
The origin of these properties is the color-ordering of the gluon fields along the path of the Wilson line, that results in a specific nested-commutator structure for exponentiated operators \cite{Vladimirov:2014wga,Vladimirov:2015fea}.
The regularization in \eq{eq:reg} definitely preserves the ordering, while naive $\delta$-regularization (with $i0$ directly replaced by $i\delta$) corresponds to some involved operator with mixed color-ordering, and violates exponentiation properties.
On the level of Feynman graphs the problems of naive $\delta$-regularization can be already seen at NNLO, where terms with color factors $C_F^2$ arise in the exponent.
Concluding, within the modified $\delta$-regularization, only diagrams with non-abelian color prefactor (\emph{web} diagrams) arise in the exponent.
Thus, the result is conveniently presented in the form
\begin{eqnarray}\label{eq:S=exp(S1+S2)}
\tilde S(b_T)=\exp\[a_s C_F\(S^{[1]}+a_sS^{[2]}+...\)\]
\,,
\end{eqnarray}
where $a_s=g^2/(4\pi)^2$ is the strong coupling and $C_F$ is the Casimir of the fundamental representation of gauge group $\(C_F=(N^2_c-1)/N_c\text{~for~}SU(N_c)\)$.

The Wilson lines in the modified $\delta$-regularization in \eq{eq:reg} do not have the gauge properties of the original Wilson lines.
However, the gauge transformation properties are restored in the limit $\d\rightarrow 0^+$. 
Therefore, only the calculation in this limit is appropriate.
In the calculation of Feynman diagrams this implies that terms linear (and higher powers) in $\delta$ should be neglected.
In this case the gauge invariance of the final result is guarantied.

Important to mention is that a naive implementation of the $\d$-regulator can cause the appearance of spurious terms that would pretend to violate gauge invariance.
This situation takes place in some diagrams at two-loop level, and in Appendix~\ref{sec:gauge} we discuss how these terms should be properly handled, and thus gauge invariance recovered.

Now, once we have introduced both the object we want to calculate and the regularization scheme we implement, we comment on the general structure of the result.
For a generic two-loop diagram, dimensional analysis gives
\begin{align}\label{eq:gen_struct}
\text{Diagram} &= \mu^{4\epsilon}\Big(A_0\pmb \delta^{-2\epsilon}+A_1\pmb \delta^{-\epsilon}\pmb B^{\epsilon}+A_2\pmb B^{2\epsilon}\Big)
+\mathcal{O}(\delta)
\,,
\end{align}
where
$$\pmb \delta=\pm \delta^+\delta^-,\quad \pmb B=\frac{b_T^2}{4}\,.$$
The sign of $\pmb \delta$ depends on the kinematics of the SF (as it can be deduced from \eq{eq:reg}): it is plus for Drell-Yan and $e^+e^-$-annihilation processes, and minus for SIDIS.
When all diagrams are summed up, the final result is linear in $\ln|\pmb\d |$ at all orders in $\epsilon$-expansion.
The linearity in $\ln|\pmb\d |$ ensures the splitting of the SF in two pieces as in \eq{eq:splitting}, and consequently the cancellation of rapidity divergences between the SF and the collinear matrix element inside any given well-defined TMD (and,
therefore, the validity of the factorization theorem as well).
Moreover, the modulus $|\pmb\d |$ makes this cancellation independent of the kinematics, being a manifestation of the universality of the SF.

The logarithms generated by rapidity divergences are partially included in the coefficient $A_2$, which is a polynomial of $\ln(\pmb \delta \pmb B)$.
The coefficients $A_0$ and $A_1$ are just functions of $\epsilon$.
Both $A_0$ and $A_1$ terms cancel exactly in the sum of diagrams.
The cancellation of the sum of the $A_0$ terms comes from the trivial fact that the integrated SF ($\pmb B=0$) is null at all positive orders in the perturbative expansion.
Thus it would not be necessary to calculate any completely virtual diagram, since they contribute only to $A_0$.
However we calculate them and explicitly check their exact cancellation.
The cancellation of $A_1$ term has a similar origin.
In fact, the integrated SF should also be recovered by taking the particular limit $\pmb B\rightarrow 0$ with $\pmb B/\pmb \d$ fixed.
Then, the only way to recover the integrated SF is to set $A_1=0$ in the sum of all diagrams.
Thus, the final result does not contain any fractional powers of $\pmb \delta$.
In this way, the only $\delta$-dependance in the final result is logarithms $\ln(\pmb \delta \pmb B)$ appearing in $A_2$.
Notice that this cancellation is also expected from the formulation of the relevant factorization theorems, where the SF is implied to be linear in the logarithms generated by rapidity divergences to all orders in $\epsilon$, and hence cannot contain $\pmb \delta^{-\epsilon}$ terms.
The rapidity divergences appear in $A_2$ only as $\ln^n(\pmb \delta \pmb B)$ with $n=1,2$, but $\ln^2(\pmb \delta \pmb B)$ cancels in the sum of all diagrams.

At higher loops we expect a similar structure of divergences in each diagram: at order $n$ the general expression for a diagram reads
$\m^{2n\e}\sum_{k=0}^{n} A_k \pmb \d^{-(n-k)\epsilon} \pmb B^{k\epsilon}$.
Similar considerations show that in the sum of all diagrams only the term $\m^{2n\e}A_n \pmb B^{n\epsilon}$ survives.
Diagram-by-diagram, the coefficient $A_n$ can contain powers of $\ln(\pmb \delta \pmb B)$ up to order $n$, while in the sum of all diagrams only terms linear in $\ln|\pmb \d|$ remain.

Finally, we observe that in order to smoothly recover the integrated SF limit with $\pmb B\rightarrow 0$, for all terms of the type $\pmb B^{n\epsilon}$ we need $\epsilon>0$.
That is also supported by the loop integrals.
In this way, all $\epsilon$-divergences of the final result are of UV origin.

We recall  that there are two sources for the UV $\epsilon$-poles of the final result.
One is the standard surface UV divergence of loop integrals, i.e. when $k^2\to \infty$.
Another one is the rapidity divergence $k^\pm \to \infty$ at fixed $k^2$.
We stress that the rapidity divergences appear both as UV and  IR divergences.
From the UV side the rapidity divergence is regularized by dimensional regularization, while from the IR side it is regularized by the $\delta$-regularization.
In the final result, the rapidity divergences appear as both $\ln(\delta)$ and $\epsilon$-poles.

Let us summarize the singularity structure of the diagrams contributing to the SF at two-loop level:
\begin{itemize}
\item
The poles in $\epsilon$ can have  at  most   a power $\epsilon^{-4}$.
Due to the renormalization theorem for Wilson lines \cite{Dotsenko:1979wb,Brandt:1981kf}, the  $\epsilon^{-4}$ poles cancel in the sum of all diagrams.
Therefore, the final result has at most $\epsilon^{-3}$ poles.
\item
In each diagram there are logarithms of $\pmb \delta$ arising from the expansion of $\pmb\delta^{-\epsilon}$ (corresponding to  the terms $A_0$ and $A_1$ in~\eq{eq:gen_struct}).
Up to order ${\cal O}(\epsilon^0)$ one can have at most $\ln^4 \pmb \delta$ contribution.
These divergences cancel in the combination of all diagrams at all orders in $\epsilon$.
\item
Finally, there are logarithms $\ln(\pmb \delta\pmb B)$ (only inside the coefficient $A_2$) and in each diagram one can have at most $\ln^2(\pmb \delta\pmb B)$.
The logarithms squared cancel in the sum of diagrams.
The remaining single logarithms are the only dependance on $\delta$ in the final result.
\end{itemize}

\section{Soft function at NLO and NNLO}
\label{sec:results}

The evaluation of the diagrams for the SF is a delicate process. One should pay a lot of attention to the analytical properties of loop
integrals, since any mistake in the $i0$-prescription gives rise to phase factors that, multiplied by $\epsilon^{-4}$, result into a wrong
combination of singularities. Mistakes of such type are very difficult to trace, and they would result into a breakdown of the TMD
factorization. Thus, we make an evaluation of all diagrams in the form shown in \eq{eq:gen_struct}, without any expansion in $\epsilon$, in
order to trace all factors and singularities explicitly. Then, combining diagrams by sectors we check that terms with $\pmb\delta^{-\epsilon}$
completely cancel with each other. This provides a very strong check of our loop integrals and for the final result.

\subsection{Soft function at NLO}

\begin{figure}[t]
\begin{center}
\includegraphics[width=0.39\textwidth]{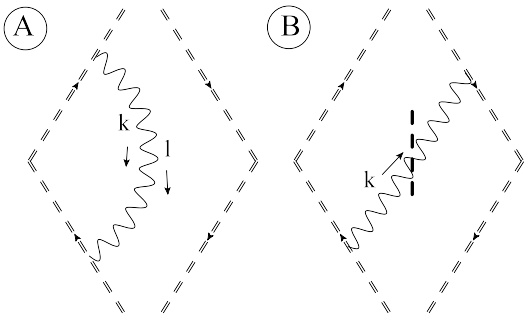}
\end{center}
\caption{\it One-loop diagrams contributing to the soft factor. The conjugated and mirror diagrams should be added.} \label{fig:1loop}
\end{figure}

The NLO diagrams are shown in Fig.~\ref{fig:1loop}, and the result reads
\begin{eqnarray}
\text{S}_A^{[1]}&=&-\frac{2g^2}{(4\pi)^{\frac{d}{2}}}\m^{2\e}\pmb\delta^{-\epsilon}\Gamma^2(\epsilon)\Gamma(1-\epsilon)
\,,
\\
\label{eq:SFB}
\text{S}_B^{[1]}&=&\frac{2g^2}{(4\pi)^\frac{d}{2}}\m^{2\e}\Big[\pmb\delta^{-\epsilon}\Gamma^2(\epsilon)\Gamma(1-\epsilon)
\nn\\&&-\pmb
B^{\epsilon}\Gamma(-\epsilon) \(L_+-\psi(-\epsilon)-\gamma_E\)\Big]
\,,
\end{eqnarray}
where
$$
L_\pm=\ln\(\pm\frac{\pmb B |\pmb \delta |}{e^{-2\gamma_E}}\)
\,,
$$
and the suffixes $A,\,B$ refer to diagrams in Fig.~\ref{fig:1loop}.
In these expressions the general structure discussed in the previous section is transparent.

Adding the mirror diagram of diagram $B$ and the complex conjugated diagram to diagram $A$ we finally obtain the complete NLO result for the SF:
\begin{eqnarray}
S^{[1]}=-4\m^{2\e}\pmb B^\epsilon\Gamma(-\epsilon)\(L_0-\psi(-\epsilon)-\gamma_E\)
\,,
\label{eq:sf1}
\end{eqnarray}
where $\psi$ stands for the digamma function and
\begin{align}
L_0=\ln\(\frac{\pmb B |\pmb \delta|}{e^{-2\gamma_E}}\)
\,.
\end{align}
The result in \eq{eq:sf1} fulfils all basic properties of the SF. 
Indeed, it is linear in $L_0$ at all orders in $\epsilon$ and thus it depends solely on $|\pmb \delta|$.

Evaluation of the one-loop diagrams clearly illustrates the structure of SF divergences discussed in the previous section. So, to evaluate the
diagram $A$ we keep $\epsilon>0$, that allows us to interpret resulting double epsilon pole as UV pole. Simultaneously, the double UV pole
appearing in diagram $A$ has an entwined structure. This double pole is collected from two sources: the surface UV singularity (that comes from
the integral over $k_T$) and the rapidity divergence (that come from the integral over $k^\pm$). 
Meanwhile, diagram $B$ is not $\epsilon$-divergent, and can be evaluated with both positive and negative $\epsilon$. 
To keep the uniform scheme we assume $\epsilon>0$, which also supports the smooth limit $\pmb B\to 0$. 
Relying on the positivity of $\epsilon$, we interpret the remaining $\epsilon$-poles in the sum of diagrams in \eq{eq:sf1} as UV poles. 
The same conclusion has been made in~\cite{Echevarria:2013aca}, where extra gluon-mass regulator was used for more detailed control of IR divergences. 
A similar structure, but with extra complications from interference of IR and UV regions,
holds as well for NNLO result.

Expanding   \eq{eq:sf1}  in $\epsilon$ we obtain (we use the standard $\overline{\rm{MS}}$-prescription  $\m^2\to\m^2e^{\g_E}/(4\pi)$)
\begin{align}
\label{eq:SF1L}
S^{[1]}=-\frac{4}{\epsilon^2}+2\mathbf{L}_\mu^2 -
\frac{2 d^{(1,1)}}{C_F} \( \frac{1} {\epsilon}+\mathbf{L}_\mu\) \mathbf{l}_\delta
+\frac{\pi^2}{3}+\mathcal{O}(\epsilon),
\end{align}
where $\mathbf{l}_\delta=\ln\(\mu^2/|\pmb \d |\)$ and $d^{(1,1)}=2 C_F=\G_0/2$.
This expression agrees with previous calculations~\cite{GarciaEchevarria:2011rb,Echevarria:2014rua,Vladimirov:2014aja,Echevarria:2015uaa}.
Notice that for the NNLO calculation we need $S^{[1]}$  up to order $\epsilon^3$ which can be derived directly from  \eq{eq:sf1}.
The coefficient $d^{(1,1)}$ refers to the quark TMD evolution $D$ function at one loop. 
We elaborate on this at all orders in Sec.~\ref{sec:SFp}.

\subsection{Soft function at NNLO}

\begin{figure*}[t]
\begin{center}
\includegraphics[width=0.85\textwidth]{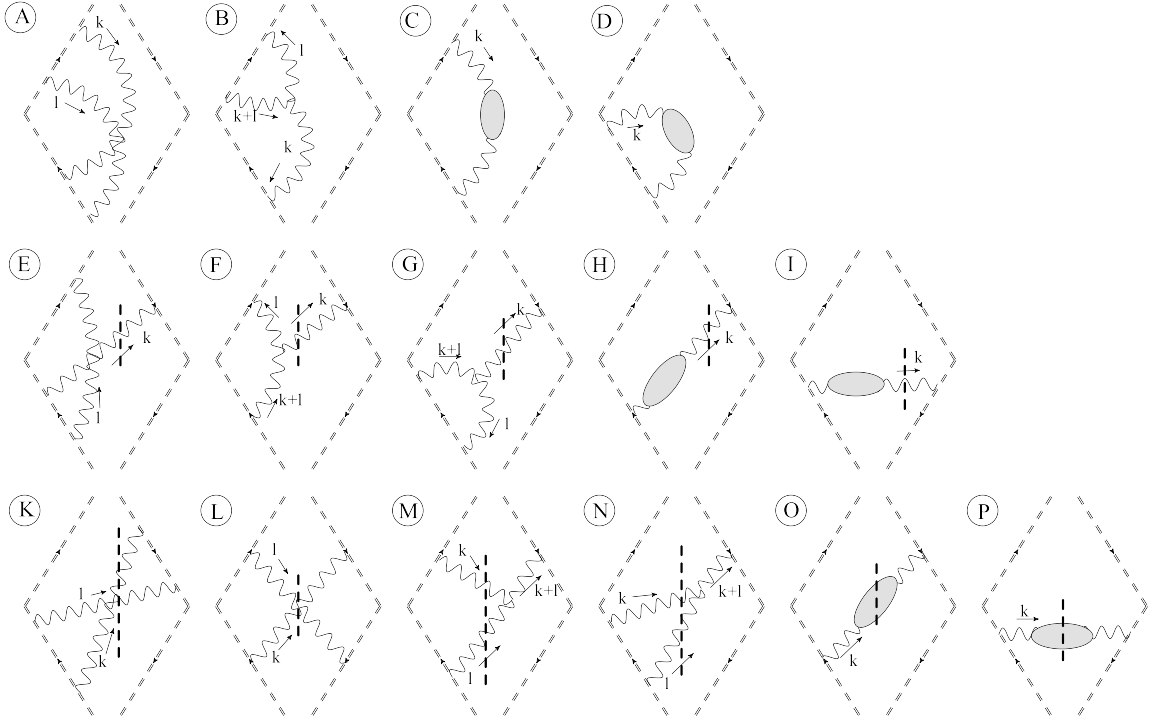}
\end{center}
\caption{\it Two-loop diagrams contributing to the soft factor grouped in virtual-virtual (VV, diagrams $A$-$D$), virtual-real (VR, diagrams
$E$-$I$), and real-real (RR, diagrams $K$-$P$). The conjugated and mirror diagrams should be added. The gray blob denotes the
vacuum-polarization subdiagram.} \label{fig:2loop}
\end{figure*}

The two-loop diagrams are shown in Fig.~\ref{fig:2loop}.
In this figure they have been grouped in virtual-virtual (VV, diagrams $A$-$D$), virtual-real (VR, diagrams $E$-$I$) and real-real (RR, diagrams $K$-$P$).
Below we discuss the main results and features of the calculation, while the technical details of the evaluation of these diagrams can be found in Appendix~\ref{appendix}.
We recall that VV diagrams contribute only to the $A_0$ terms in \eq{eq:gen_struct}, so they cancel completely in the sum of all diagrams.

Besides ordering the diagrams in VV, VR and RR, the two-loop web diagrams can be naturally split into four categories:
\begin{itemize}
\item[i)]
Diagrams with vacuum-polarization subdiagrams ($C$, $D$, $H$, $I$, $O$, $P$);
\item[ii)]
Multi-gluon exchange webs (MGEW) ($A$, $E$, $K$, $L$);
\item[iii)]
Diagrams with 3-gluon vertex ($B$, $F$, $G$, $M$, $N$);
\item[iv)]
Counterterms that come from the renormalization of the Wilson lines.
\end{itemize}
These subclasses represent sectors of QCD interactions with different properties.
Therefore, it is convenient to present the expression for the SF in the form
\begin{eqnarray}
\label{eq:scomb}
S^{[2]}=S^{[2]}_{pol}-\frac{C_A}{2} S^{[2]}_{\text{MGEW}}-\frac{C_A}{2}S^{[2]}_{3g}+S^{[2]}_{ct}
\,.
\end{eqnarray}
The mixed singularities proportional to $\pmb\delta^{-\epsilon}$ cancel inside every class of diagrams. 
This happens because these classes represent separate parts of QCD interactions, while the cancellation is dictated by the structure of the operator. 
The confirmation of the cancellation inside every class of diagrams is a very strong check of the calculation. Another check is the cancellation of the leading
$\epsilon^{-4}$-pole. Due to the QCD Ward identities the cancellation takes place between MGEW and 3-gluon interaction diagrams.

The result of the contribution of the diagrams with a  vacuum polarization subdiagram  is
\begin{align}
\label{eq:Spol}
S^{[2]}_{pol}&=8\m^{4\e}\pmb B^{2\epsilon}\[C_A (5-3\epsilon)-4T_RN_f(1-\epsilon)\]
\\
&
\qquad\nn
\times\frac{\Gamma(-2\epsilon)\Gamma(-\epsilon)\Gamma(2-\epsilon)}{\Gamma(4-2\epsilon)}
\(L_0-\psi(-2\epsilon)-\gamma_E\)
\\
&\nn
-8\m^{2\e}\pmb B^\epsilon \(\frac{2}{3}T_RN_f-\frac{5}{6}C_A\)\frac{\Gamma(-\epsilon)}{\epsilon}\(L_0-\psi(-\epsilon)-\gamma_E\)
\,.
\end{align}
The term proportional to $\pmb B^{2\epsilon}$ in \eq{eq:Spol} comes entirely from diagram $O$. The part of diagram $O$ proportional to $\pmb
\delta^{\epsilon}$ cancels with similar contributions coming from other diagrams (see explicit expressions in Appendix \ref{appendix}).
Diagrams $H$ and $I$ are zero in dimensional regularization, however their UV counterterms are not. The last line in \eq{eq:Spol} is the
contribution of the UV renormalization of diagram $H$, and it is proportional to the one-loop diagram $B$, \eq{eq:SFB}, multiplied by
gluon propagator renormalization constant $Z_{3,\overline{MS}}^{1/2}$ ,
\begin{align}
\label{eq:z3} Z_{3,\overline{MS}}=1-\frac{a_s}{3\epsilon}\left(4 T_R N_f-5 C_A\right)+\mathcal{O}(a_s^2).
\end{align}
Notice  that $S^{[2]}_{pol}$ is linear in $L_0$.

The contribution of MGEW diagrams collected together is
\begin{widetext}
\begin{align}\label{eq:MGEW}
S^{[2]}_{\text{MGEW}}&=4\m^{4\e}\pmb B^{2\epsilon}
\Bigg\{\Gamma^2(-\epsilon)\(L_0-\psi(-\epsilon)-\gamma_E\)^2 +4
\Gamma(-2\epsilon)\Gamma(-\epsilon)\Gamma(\epsilon)\(\psi\(\frac{1-\epsilon}{2}\)-\Gamma(-\epsilon)\Gamma(1+\epsilon)\)+4 Q(\epsilon)
\\&\nn+\Gamma^2(-\epsilon)\[\(2L_0+2\gamma_E+8\ln
2-\psi(-2\epsilon)-3\psi(-\epsilon)\)\(\psi(-\epsilon)-\psi(-2\epsilon)\)+3\psi'(-2\epsilon)-2\psi'(-\epsilon)-\frac{5\pi^2}{6}\]\Bigg\},
\end{align}
where the function $Q(\epsilon)=\mathcal{O}(\epsilon^0)$ and it is given in \eq{app:def_Q}.

The result for the  3-gluon interaction diagrams  is
\begin{eqnarray}\label{eq:3g}
S^{[2]}_{3g}&=&-4\m^{4\e}\pmb B^{2\epsilon}\Gamma^2(-\epsilon)\Bigg\{\(L_0-\psi(-\epsilon)-\gamma_E\)^2 \\&& \nn
+2\Big[\(L_0-\psi(-2\epsilon)-\gamma_E\)\(\frac{1}{1-2\epsilon}+\psi(-2\epsilon)-\psi(1-\epsilon)+\psi(1+\epsilon)+\gamma_E\)
+\frac{\ln2}{1-2\epsilon}-\frac{\pi^2}{6}
\\&& \nn +\psi'(-\epsilon)+\frac{1}{2}\psi'(1+\epsilon)-\frac{3}{2}\psi'(-2\epsilon)
-\frac{1}{2}\(\psi(-\epsilon)+\gamma_E\)\(2\psi(-2\epsilon)-3\psi(-\epsilon)-\gamma_E\)
\\&& \nn +\frac{1}{2}\(\psi(-2\epsilon)+\psi(1+\epsilon)+2\gamma_E\)\(3\psi(-2\epsilon)-4\psi(-\epsilon)+\psi(1+\epsilon)\)
+\frac{\psi(-2\epsilon)-\psi(-\epsilon)}{\epsilon}-\frac{1}{2\epsilon^2}\Big]\Bigg\}.
\end{eqnarray}
\end{widetext}

The logarithms $L_0$ appear quadratically and linearly in both \eq{eq:MGEW} and \eq{eq:3g}.
However in their sum the $L_0^2$ terms cancel with each other, leaving a linear dependence on $L_0$.
It is worth mentioning that this cancellation leaves a trace in the form of terms proportional to $\pi^2$.
It happens because, individually, diagrams depend on $L_\pm$ and their complex conjugates.
In the final result, the complex phases cancel and~\eq{eq:MGEW} and~\eq{eq:3g} are
naturally expressed via $L_0$.

Finally, we should add the renormalization factor for Wilson lines, that comes from the renormalization of both the coupling constant and the gluon fields.
It is proportional to the one-loop diagrams multiplied by $Z_{g\overline{MS}} Z_{3\overline{MS}}^{1/2}$.
$Z_{3\overline{MS}}$ is given in \eq{eq:z3}, and
\begin{align}
Z_{g\overline{MS}}=1-\frac{a_s}{6 \epsilon}\left( 11 C_A-4 T_R N_f\right)+\mathcal{O}(a_s^2)
\,,
\end{align}
which add up to
\begin{eqnarray}
S^{[2]}_{ct}=8a_s^2C_FC_A\frac{\Gamma(-\epsilon)}{\epsilon}\m^{2\e}\pmb B^\epsilon\(L_0-\gamma_E-\psi(-\epsilon)\).
\,.
\end{eqnarray}

Combining all pieces together as in \eq{eq:scomb}, we obtain the final expression for the SF.
Here we explicitly write the result expanded in $\epsilon$:
\begin{widetext}
\begin{eqnarray}
\label{eq:Stot}
S^{[2]}&=&\frac{d^{(2,2)}}{C_F}\(\frac{3}{\epsilon^3}+\frac{2\mathbf{l}_\delta}{\epsilon^2}+\frac{\pi^2}{6\epsilon}+\frac{4}{3}\mathbf{L}_\mu^3-2\mathbf{L}_\mu^2\mathbf{l}_\delta
+ \frac{2\pi^2}{3}\mathbf{L}_\mu+\frac{14}{3}\zeta_3\)
-\frac{d^{(2,1)}}{C_F}\(\frac{1}{2\epsilon^2}+\frac{\mathbf{l}_\delta}{\epsilon}-\mathbf{L}_\mu^2+2\mathbf{L}_\mu\mathbf{l}_\delta-\frac{\pi^2}{4}\)
\nn\\&&
-\frac{d^{(2,0)}}{C_F}\(\frac{1}{\epsilon}+2\mathbf{l}_\delta\)+C_A\(\frac{\pi^2}{3}+4\,\ln 2\)\(\frac{1}{\epsilon^2}+\frac{2\mathbf{L}_\mu}{\epsilon}
+2\mathbf{L}_\mu^2+\frac{\pi^2}{6}\)+C_A\(8\,\ln 2-9\zeta_3\)\(\frac{1}{\epsilon}+2\mathbf{L}_\mu\)
\nn\\
&&
+\frac{656}{81}T_RN_f+C_A\(-\frac{2428}{81}+16\,\ln2-\frac{7\pi^4}{18}-28\,\ln
2\,\zeta_3+\frac{4}{3}\,\pi^2\ln^22-\frac{4}{3}\,\ln^42-32\text{Li}_4\(\frac{1}{2}\)\)+\mathcal{O}(\epsilon)
\,.
\end{eqnarray}
\end{widetext}
The coefficients  $d^{(n,k)}$ are the coefficients of the  D-function  at two loops which governs
TMD  evolution kernel (see~\cite{Echevarria:2012pw}, which uses a slight different notation),
\begin{eqnarray}\nn
d^{(2,2)}&=&\frac{\Gamma^{(0)}\beta_0}{4}=C_F\(\frac{11}{3}C_A-\frac{4}{3}T_RN_f\)
\,,
\\\label{eq:d_AD}
d^{(2,1)}&=&\frac{\Gamma^{(1)}}{2}=2C_F\!\!\(\!\!\(\frac{67}{9}-\frac{\pi^2}{3}\)C_A-\frac{20}{9}T_RN_f\)
\,,
\nn\\
d^{(2,0)}&=&C_F\(\(\frac{404}{27}-14\zeta_3\)C_A-\frac{112}{27}T_RN_f\).
\end{eqnarray}
We discuss the relation between the SF and the $D$ function more extensively in next Section.

The final result in \eq{eq:Stot} contains terms with $\ln 2$, which are atypical in loop-calculations.
This is a feature of the $\delta$-regulator.
However, these terms cancel within an object free from rapidity divergences (see the result for the TMD fragmentation function obtained in~\cite{Echevarria:2015usa}).

\section{Extraction of the evolution factor $D$ from the SF}
\label{sec:SFp}

The evolution factor $D$ for quark TMDs can be extracted directly from the SF. 
This feature is fundamental to prove the universality of the TMD evolution. 
In fact, because of the universality (or process independence) of the SF, the function $D$ is also universal and appears in the evolution of all (un)polarized TMDs, both distribution
and fragmentation functions.

To extract the evolution factor $D$, we recall the definition of a generic TMD in the light-cone $+$ direction
\begin{align}\label{eq:Tefinition}
T_{+}(z, b_T;\zeta_+,\m) &= Z_2^{-1}\, Z_T\(\frac{\zeta_+}{\mu^2}\)\, T^{unsub}_+\(z, b_T;\mu,\frac{\delta^+}{p^+}\)
\nn\\
&\times \tilde S^{-1/2}\(b_T;\mu,\d^+/\n\) 
\,,
\end{align}
where $Z_2$ is the quark wave-function renormalization constant, $Z_T$ the TMD operator renormalization constant and $\tilde S$ is evaluated in the previous Section. 
The variable $\zeta_+ = (p^+/\n)^2$ is the rapidity scale that arises from the freedom in the spliting procedure of the SF (see e.g. \cite{Echevarria:2012pw,Echevarria:2015usa} for more details), with $\nu$ an arbitrary and positive real number that transforms as $p^+$ under boosts (cfr.~\eq{eq:splitting})

The dependence on $\d^+$ is enclosed in the SF and
the unsubtracted TMD, while the rapidity scale variation is dictated  by the TMD renormalization factor $Z_T$ and the SF.

The TMD evolution equation with respect to rapidity parameter defines the evolution factor $D$. It reads
\begin{align}
\label{eq:Tevolution} \frac{d}{d\ln\zeta_+} T_+&=-D \;T_+
\,.
\end{align}
Only the SF and the renormalization constant $Z_T$ are $\zeta$-dependent. 
Therefore, we rewrite
\begin{eqnarray}
\frac{d \ln Z_T\(\frac{\zeta_+}{\mu^2}\)}{d\ln\zeta_+}
- \frac{1}{2}\frac{d \ln \tilde S\(\d^+\frac{\sqrt{\z_+}}{p^+}\)}{d\ln \zeta_+}
= -D
\,,
\end{eqnarray}
where we drop unnecessary arguments of functions. 
Finally, using the relation in \eq{eq:splitting} we obtain
\begin{eqnarray}\label{eq:D_fromS}
D=\frac{1}{2}\frac{d \ln \tilde S}{d\mathbf{l}_\delta}-\frac{d \ln Z_T}{d\ln\zeta_+}.
\end{eqnarray}

Evaluating the derivative of the SF we obtain a result that contains $\epsilon$-poles. 
However, these $\epsilon$-poles are removed by the renormalization constant $Z_T$ in \eq{eq:D_fromS}, so the function $D$ is finite. 
It is important that renormalization constant $Z_T$ contains only $\epsilon$-poles and does not contain finite $\ln(\zeta)$ dependance. 
Therefore, we can extract the function $D$ by differentiating the SF and neglecting all $\epsilon$-poles. 
The perturbative result for the $D$-function can be written as
\begin{align}
D&=\sum_{n}\sum_{k=0}^n a_s^n \pmb L_\m^k d^{(n,k)},
\end{align}
the NLO coefficients are: $d^{(1,0)}=0$, $d^{(1,1)}$ is given after \eq{eq:SF1L}, while NNLO coefficients $d^{(2,i)}$ are given in \eq{eq:d_AD}.
The calculation of the SF performed  in this  work so represents  the first direct calculation of the  $D$-function at NNLO and agrees with
previous  derivation of the result~\cite{Becher:2010tm,GarciaEchevarria:2011rb}.

We conclude noting that a remarkable property of non-abelian exponentiation is that the SU(3) generators in the Wilson lines enter as a global multiplier of the exponent, see ~\eq{eq:SF_def2}. 
In this way, the difference between soft functions that use different SU(3) representations is a multiplicative factor at all orders in perturbation theory.
This implies that the $D$-function for gluons, $D_g$, and quarks, $D$, are related by
\begin{align}
\label{eq:scaling} \frac{D}{C_F}=\frac{D_g}{C_A}
\end{align}
at all orders in perturbation theory.

\section{Conclusions}
\label{sec:conclusions}

The definition and calculation of the soft function is crucial in TMD factorization theorems.
In fact, one has just one soft function for all quark TMDs (and one for all gluon TMDs), which
enters their definition in a specific way to make them well-defined hadronic quantities.
This relies on the single logarithmic dependence of the soft function in the logarithm generated by rapidity divergences, which allows its splitting in rapidity space.
In this work we have reported on the NNLO calculation of this soft function, using the $\d$-regulator for rapidity divergences, and explicitly checked for the first time its splitting in rapidity space at this non-trivial order.
Moreover, we have performed the calculation to all orders in dimensional regularization, which allows the extraction of the general structure of the soft function to all orders.

From the result we have obtained the $D$ function at NNLO, which represents the first independent and direct calculation of this universal quantity, relevant for the evolution of TMDs.
Furthermore, given that the relation between the quark and gluon soft functions is just a  Casimir scaling at all orders in perturbation theory, we have also obtained the NNLO soft function relevant for gluon TMDs, as well as its $D_g$ term at the same order.

Finally, since the soft function is universal and enters the definition of all (un)polarized TMD distribution and fragmentation functions, this calculation represents an important ingredient in order to obtain the relevant perturbative coefficients to resum large logarithms for any TMD.
A first application of this soft function which represents also a strong check of the present calculation is reported in~\cite{Echevarria:2015usa}.

\begin{acknowledgments}
A.V. thanks Victor Svensson for helpful discussions. M.G.E. is supported by the ``Stichting voor Fundamenteel
Onderzoek der Materie'' (FOM), which is financially supported by the ``Nederlandse Organisatie voor Wetenschappelijk Onderzoek'' (NWO). I.S. is
supported by the Spanish MECD grants FPA2011-27853-CO2-02 and FPA2014-53375-C2-2-P.
\end{acknowledgments}

\appendix
\section{Expressions for diagrams and loop integrals}
\label{appendix}

In this Appendix we present the expressions for individual diagrams, which are shown in Figs.~\ref{fig:1loop} and~\ref{fig:2loop}.
We use the notation $S^{[n]}_{X}$ for the contribution of diagram $X$ to $S^{[n]}$, defined in \eq{eq:S=exp(S1+S2)}.

The calculation of the loop-integrals was performed in the following way.
First, we integrate over one of the light-cone components, using either the $\delta$-function for real gluons or Cauchy theorem for virtual components.
Second, we integrate over transverse components.
Here for real gluons we make the Mellin-Barnes expansion, such that the integrand contains only powers of the transverse components, such that the Fourier integral can be easily performed.
Third, we integrate over the residual light-cone component.
Finally, we are left with an integral of Mellin-Barnes type, that can be straightforwardly evaluated by closing the integration contour in the half-plane with suppressed $\pmb\delta$.
Typically, one needs to consider only the residues in the vicinity of zero, since we need only the leading terms in $\pmb\delta\to 0$.
However, for some integrals (e.g. $I''_A$) powers of $\pmb \delta$ can be compensated between the Mellin-Barnes integrals, and the complete sum of residues should be considered.

The one-loop diagrams are
\begin{eqnarray}
S^{[1]}_{A}&=&-\m^{2\e}K_1^{(0)}
\,,
\\\nn
S^{[1]}_B&=&-\m^{2\e}K_1'
\,.
\end{eqnarray}
The two-loop diagrams with two virtual gluons are
\begin{eqnarray}\nn
S^{[2]}_A&=&-\frac{C_A}{2}\m^{4\e}I_A,
\\
S^{[2]}_B&=&-\frac{C_A}{2}\m^{4\e}(2I_{C1}+I_{C2}),\label{eq:ABC}
\\\nn
S^{[2]}_C&=&-4\Gamma(\epsilon)\frac{\Gamma(1-\epsilon)\Gamma(3-\epsilon)}{\Gamma(5-2\epsilon)}
\\&&\times\[C_A(5-3\epsilon)-4T_RN_f(1-\epsilon)\]\m^{4\e}\(K_1^{(\epsilon)}+...\),\nn
\end{eqnarray}
where the dots denote a term which is zero if the $\d$-regulator is implemented properly (see explanation in Appendix~\ref{sec:gauge}).
Diagram $D$ is zero (see also Appendix~\ref{sec:gauge}).
The two-loop diagrams with a single real gluon are
\begin{eqnarray}\nn
S^{[2]}_E&=&-\frac{C_A}{2}\m^{4\e}I_A',
\\
S^{[2]}_{F}&=&\frac{C_A}{2}\m^{4\e}(I'_{C3}-I'_{C4}),
\\\nn
S^{[2]}_{G,H,I}&=&0
\,,
\end{eqnarray}
where the latter are zero are due to the absence of a scale within the virtual loop.
Finally, the diagrams with two real gluons are
\begin{eqnarray}\nn
S^{[2]}_K&=&-\frac{C_A}{2}\m^{4\e}I''_A,
\\\nn
S^{[2]}_{L}&=&-\frac{C_A}{2}\m^{4\e}\big|K'_1\big|^2,
\\ S^{[2]}_{M}&=&\frac{C_A}{2}\m^{4\e}(I''_{C1}+I''_{C2}),\label{eq:OP}
\\\nn
S^{[2]}_{N}&=&\frac{C_A}{2}\m^{4\e}(I''_{C3}-2I''_{C4}),
\\\nn
S^{[2]}_{O}&=&-2(C_A(5-3\epsilon)-4T_RN_f(1-\epsilon))
\\\nn&&\times\frac{\Gamma(2-\epsilon)}{\Gamma(4-2\epsilon)}\m^{4\e}(K''_1+...)
\,,
\end{eqnarray}
where the dots denote a term which is zero if the $\d$-regulator is implemented properly (see explanation in Appendix~\ref{sec:gauge}).
Diagram $P$ is zero for the same reason (see also Appendix~\ref{sec:gauge}).

The expressions for one-loop-like integrals are
\begin{eqnarray}\nn
K_1^{(a)}&=& \int\!\! \frac{d^dk}{(2\pi)^d}\frac{-i}{(k^++i\delta^+)(k^-+i\delta^-)(-k^2-i0)^{1+a}}
\\\nn
&=&2\pmb\delta^{-\epsilon-a}\frac{\Gamma^2(a+\epsilon)\Gamma(1-a-\epsilon)}{\Gamma(1+a)},
\\\nn
K'_1&=&\int\frac{d^dk}{(2\pi)^d}\frac{(-2\pi)\theta(k^+)\delta(k^2)e^{i(kb)_T}}{(k^++i\delta^+)(k^-+i\delta^-)}
\\&=&-2\[\pmb\delta^{-\epsilon}\Phi_\epsilon-\pmb B^{\epsilon}\Psi_\epsilon\],
\\\nn
K''_1&=&\int\!\!
\frac{d^dk}{(2\pi)^d}\frac{(2\pi)\theta(k^+)\theta(k^2)e^{i(kb)_T}(k^2)^{1-\epsilon}}{(k^+\!+\!i\delta^+\!)(k^-\!+\!i\delta^-\!)(k^2\!+\!i0\!)(k^2\!-\!i0\!)}
\\\nn&=&
2\Gamma(-\epsilon)\[\pmb\delta^{-2\epsilon}\Phi_{2\epsilon}-\pmb B^{2\epsilon}\Psi_{2\epsilon}\],
\end{eqnarray}
where we use the short hand notation
$$
\Phi_\epsilon=\Gamma^2(\epsilon)\Gamma(1-\epsilon),
$$
$$
\Psi_\epsilon=\Gamma(-\epsilon) \(L_+-\psi(-\epsilon)-\gamma_E\).
$$
The expressions for the MGEW integrals are
\begin{widetext}
\begin{eqnarray}\nn
I_A&=&\int \frac{d^dk d^dl}{(2\pi)^{2d}}\frac{-1}{(k^+-i\delta^+)(k^++l^+-2i\delta^+)(l^--i\delta^-)(l^-+k^--2i\delta^-)(k^2+i0)(l^2+i0)}
=\frac{1}{4}\(K_1^{(0)}\)^2
\,,
\\ \nn I'_A&=&\int
\frac{d^dkd^dl}{(2\pi)^{2d}}\frac{(-2\pi i)
~\theta(k^+)\delta(k^2)e^{i(kb)_T}}{(l^++i\delta^+)(l^-+i\delta^-)(l^-+k^-+2i\delta^-)(k^++i\delta^+)(l^2+i0)}
\\
&=&K_1^{(0)}K'_1+4\Phi_\epsilon\(\frac{\pmb \delta^{-2\epsilon}}{2}\Phi_\epsilon+\pmb B^\epsilon \pmb \delta^{-\epsilon}F_\epsilon+\pmb
B^{2\epsilon}\Gamma(-2\epsilon)\Gamma(-\epsilon)\)
\,,
\\\nn
I''_A&=&\int \frac{d^dkd^dl}{(2\pi)^{2d}}\frac{(-2\pi)^2~\theta(k^+)e^{i(kb)_T}\delta(k^2)\theta(l^+)e^{i(lb)_T}\delta(l^2)}{
(l^++i\delta^+)(k^++l^++2i\delta^+)(l^-+i\delta^-)(k^-+l^-+2i\delta^-) } = \frac{(K'_1)^2}{2}-\pmb \delta^{-2\epsilon}\Phi^2_\epsilon-8\pmb
B^\epsilon \pmb \delta^{-\epsilon}\Phi_\epsilon F_\epsilon
\\\nn&&-2 \pmb B^{2\epsilon}\Psi^2_\epsilon
 +4\pmb B^{2\epsilon}\Phi_\epsilon
\(\psi(-\epsilon)-\psi(-2\epsilon)\)+8\pmb
B^{2\epsilon}\(\Gamma(-2\epsilon)\Gamma(-\epsilon)\Gamma(\epsilon)\psi\(\frac{1-\epsilon}{2}\)+Q(\epsilon)\)
\\&&\nn+2\pmb B^{2\epsilon}\Gamma^2(-\epsilon)\Big[\frac{\pi^2}{6}+3\psi'(-2\epsilon)-2\psi'(-\epsilon)+
(\psi(-2\epsilon)-\psi(-\epsilon)+4\gamma_E+8\ln2)(\psi(-\epsilon)-\psi(-2\epsilon))\Big]
\,,
\end{eqnarray}
 where
\begin{eqnarray}\nn
F_\epsilon&=&2^{1-\epsilon}\frac{\Gamma(-\epsilon)}{\epsilon}\,_2F_1(1,1,1+\epsilon;-1),
\\\label{app:def_Q}
Q(\epsilon)&=&\sum_{k=1}^\infty\frac{(-1)^k}{k!}\Gamma(k-\epsilon) \Big(\Gamma(k)\Gamma(-k-\epsilon)\psi\(\frac{1+k}{2}\)
+\Gamma(k-2\epsilon)\Gamma(\epsilon-k)\psi\(\frac{1+k-\epsilon}{2}\)\Big)
\,.
\end{eqnarray}
The integrals involving three-gluon vertex are
\begin{eqnarray}\nn
I_{C1}&=&\int \frac{d^dkd^dl}{(2\pi)^{2d}}\frac{1}{(k^++i\delta^+)(k^-+2i\delta^-)(k^2+i0)(l^2+i0)[(k+l)^2+i0]}
=2^{-2\epsilon}\Gamma(\epsilon)\frac{\Gamma^2(1-\epsilon)}{\Gamma(2-2\epsilon)}K_1^{(\epsilon)},
\\\nn
I_{C2}&=& \int \frac{d^dkd^dl}{(2\pi)^{2d}}\frac{1}{(k^++i\delta^+)(l^--i\delta^-)(k^2+i0)(l^2+i0)[(k+l)^2+i0]}=
 2\pmb
\delta^{-2\epsilon}\Phi_{2\epsilon}\Gamma(\epsilon)\Gamma(-\epsilon)
\\\nn I'_{C3}&=&\int
\frac{d^dkd^dl}{(2\pi)^{2d}}\frac{(-2\pi i)~\theta(k^+)\delta(k^2)e^{i(kb)_T}}{(l^++i\delta^+)(k^-+l^-+i\delta^-)(l^2+i0)((k+l)^2+i0)}
=
-I''_{C1}-2I_{C2}-4\pmb B^{\epsilon}\pmb\delta^{-\epsilon}\Gamma^2(\epsilon)\Gamma^2(-\epsilon)
\\&&+\pmb B^\epsilon\Gamma^2(-\epsilon)\Big[(L_++\psi(-2\epsilon)+\psi(1+\epsilon)-2\psi(-\epsilon))^2
+2\psi'(-\epsilon)+\psi'(1+\epsilon)-3\psi'(-2\epsilon)+\frac{2\pi^2}{3}\Big]
\\\nn I'_{C4}&=&\int
\frac{d^dkd^dl}{(2\pi)^{2d}}\frac{(-2\pi i)~\theta(k^+)\delta(k^2)e^{i(kb)_T}}{(k^++i\delta^+)(k^-+l^-+i\delta^-)(l^2+i0)((k+l)^2+i0)}=0,
\\\nn I''_{C1}&=& \int
\frac{d^dkd^dl}{(2\pi)^{2d}}\frac{(-2\pi)^2~\theta(k^+)\delta(k^2)e^{i(kb)_T}~\theta(l^+)\delta(l^2)e^{i(lb)_T}}{(k^+-i\delta^+)(l^-+i\delta^-)((k+l)^2-i0)}
\\\nn &=&
2(-\pmb \delta)^{-2\epsilon}\Phi_{2\epsilon}\Gamma(\epsilon)\Gamma(-\epsilon)+ \pmb
B^{2\epsilon}\Gamma^2(-\epsilon)\Big[(L_0+\psi(-2\epsilon)-\psi(-\epsilon))^2-3\psi'(-2\epsilon)+\psi'(-\epsilon)+\frac{2\pi^2}{3}\Big]
\\\nn I''_{C2}&=&I''_{C3}=
\int
\frac{d^dkd^dl}{(2\pi)^{2d}}\frac{(-2\pi)^2~\theta(k^+)\delta(k^2)e^{i(kb)_T}~\theta(l^+)\delta(l^2)e^{i(lb)_T}}{(k^++l^++i\delta^+)(l^-+i\delta^-)((k+l)^2-i0)}
\\\nn &=& 2\pmb \delta^{-2\epsilon}\Phi_{2\epsilon}\Gamma(\epsilon)\Gamma(-\epsilon) +2\pmb B^{\epsilon}\pmb
\delta^{-\epsilon}\Gamma^2(\epsilon)\Gamma^2(-\epsilon) +\pmb
B^{2\epsilon}\(\frac{\Psi_\epsilon}{\epsilon}-\frac{\Gamma^2(-\epsilon)}{\epsilon}\(\psi(-2\epsilon)-\psi(1-2\epsilon)\)\)
\\ \nn I''_{C4}&=&
\int
\frac{d^dkd^dl}{(2\pi)^{2d}}\frac{(-2\pi)^2~\theta(k^+)\delta(k^2)e^{i(kb)_T}~\theta(l^+)\delta(l^2)e^{i(lb)_T}}{(k^++l^++i\delta^+)(k^-+l^-+2i\delta^-)((k+l)^2-i0)}
=-I''_{C1}-\pmb B^{2\epsilon}\frac{\Psi_\epsilon+\ln 2+\psi(-\epsilon)-\psi(-2\epsilon)}{1-2\epsilon}.\nn
\end{eqnarray}
\end{widetext}

The explicit form of $Q(\epsilon)$ that we found is a complicated expression that involves derivatives of $\,_3F_2$-hypergeometric function.
The $\epsilon$-expansion of $Q(\epsilon)$ reads
\begin{eqnarray}
Q(\epsilon)&=&-\frac{23\pi^4}{1440}-\frac{\pi^2}{6}\ln^22-2\gamma_E\zeta_3+\frac{\ln^42}{6} \\&&\nn+4 \text{Li}_4\(\frac{1}{2}\)-\frac{\ln
2}{2}\zeta_3+\mathcal{O}(\epsilon^1)
\,.
\end{eqnarray}

\section{Gauge invariance and $\d$-regulator}
\label{sec:gauge}

It is legitimate to study whether the $\d$-regulator that we have used interferes with the gauge invariance of the final result.  
Given the implementation of the regulator at the level of the operator in~\eq{eq:reg}, one would naively expect that eventual gauge breaking pieces appear as positive powers of the $\d$ parameter, and so disappear in the limit $\d^{\pm}\to 0$.
However in the perturbative computations we observe that a naive implementation of the $\d$-regulator breaks gauge invariance.

The problems with gauge invariance are caused by the loop-integrals which do not contain any rapidity divergences, and which should then not be calculated while keeping the $\d$'s finite.
In this regard, within our NNLO calculation, only the longitudinal ``$k^\m k^\n$-part'' of the gluon self-energy $\Pi^{\m\n}$ is worrisome, since in some of the diagrams it can cancel the rapidity divergences and allow for the integrals to be done in pure dimensional regularization, after setting all $\d$'s to zero.
In particular, these integrals appear only in diagrams $C$, $O$ and $P$.
Below we explain in detail what is the issue in diagram $C$, but similar considerations apply to $O$ and $P$.

Diagram $C$ is given by the following integral:
\begin{align}
\text{SF}_{C} &=
g^2C_F\int\frac{d^dkd^dl}{(2\pi)^{2d}}
\nn\\
&\quad\times
\frac{-\bn^\m n^\n
}{(k^++i\delta^+)(-k^--i\delta^-)}\frac{-i}{k^2}\Pi_R^{\mu\nu}(k)\frac{-i}{k^2}
\,,
\end{align}
where $\Pi_R^{\m\n}$ is the renormalized gluon self-energy. 
It reads
\begin{align}
\Pi^{\m\n}_R(k) &=
ia_s\frac{(-1)^{-\e}}{(k^2+i0)^\e}\(g^{\m\n}k^2-k^\m k^\n\)
\G(\e)
\nn\\
&\quad
\times
\frac{4\G(1-\e)\G(3-\e)}{\G(5-2\e)}
\[C_A(5-3\e) -T_f N_f4(1-\e)\]
\nn\\
&\quad
+i(1-Z^{MS}_3)\(g^{\m\n}k^2-k^\m k^\n\)
\,,
\end{align}
with
$$
Z^{MS}_3=1-\frac{a_s}{\epsilon}\(\frac{4}{3}T_RN_f-\frac{5}{3}C_A\)+...~.
$$
As it can be seen, the longitudinal ``$k^\m k^\n$-part'' of the gluon self-energy will give a term where $\d$'s can be set to zero at the level of the integrand, since there are no rapidity divergences to be regularized.
Thus this term can be calculated in pure dimensional regularization, giving zero.
The same logic applies to diagrams $O$ and $P$, where the longitudinal part of the self-energy sub-diagram will give terms that do not contain rapidity divergences, and which are zero in pure dimensional regularization.

Concluding, the gauge invariance of the final result is completely guaranteed as far as the $\d$-regulator is implemented consistently, i.e., respecting the axiomatic properties of dimensional regularization.
In other words, $\d$'s should be set to zero in every integral that does not contain rapidity divergences.

\end{document}